\documentclass[12pt]{article}
\usepackage{epsfig}
\usepackage{cite}

\begin{document}

\title{SLR Contribution to Investigation of Polar Motion\footnote{In: ASP Conf. Ser., v.~208, Polar Motion: Historical and Scientific
  Problems, Proc. IAU Coll. 178, Cagliari, Italy, Sep 27-30, 1999, Eds. S. Dick, D. McCarthy, B. Luzum, 2000, 267-276.}}
\author{Zinovy Malkin\footnote{Current affiliation: Pulkovo Observatory, St.~Petersburg, Russia} \\ Institute of Applied Astronomy, St.Petersburg, Russia}
\date{\vspace{-10mm}}
\maketitle

\begin{abstract}
SLR technique has being used for determination
of ERP during over twenty years.
Most of results contributed to IERS are based
on analysis of observations of Lageos 1\&2 satellites collected
at the global tracking network of about 40 stations.
Now 5 analysis centers
submit operative (with 2-15 days delay) solutions and about
10 analysis centers yearly contribute final (up to 23 years) ERP series.
Some statistics related to SLR observations and analysis is presented
and analyzed.
Possible problems in SLR observations and analysis and ways of
its solution are discussed.
\end{abstract}

\section{Introduction}

Laser ranging to Earth artificial satellites (SLR) was initiated
in 1964 after launch of the first geodetic/geodynamical satellite Beacon-B.

Since that time satellite laser ranging (SLR) technique have being
widely used for geodynamical and geophysical researches.
The primary fields of investigations used SLR observations are Earth
rotation, maintenance of the Terrestrial Reference Frame,
tectonic motion, Earth crust deformations, geopotential
with its spatial and temporal variations, tides, movement of geocenter,
support of satellite geophysical missions (such as satellite altimetry),
global time transfer, and others.
Detailed analysis of SLR contribution to Earth sciences can be found
in (Tapley {\em et al.} 1993).

After the launching of the Lageos
satellite in May 1976, SLR became one of the main techniques for
investigations of the Earth rotation, and
during over twenty years SLR technique have being used for determination
of ERP.
Most of results contributed to IERS are based
on analysis of observations of Lageos 1\&2 satellites collected
at the global tracking network of about 40 stations.
SLR provides high
precision series of Xp, Yp, and LOD. Some analysis centers compute also UT
that allow to densify Universal Time series in combination with VLBI data.

Importance of ILRS as one of the main method to study the Earth led,
naturally, to the establishment of the International Laser Ranging
Service (ILRS) in 1998 that coordinates now scientific activity
SLR (and LLR), chiefly in the framework of IAG and IERS projects.

In this paper we will focus only on SLR contribution to investigation
of polar motion in accordance with topics of the conference.

\section{Contribution of SLR PM series to the IERS}

\begin{figure}
\epsfxsize=\textwidth
\centerline{\epsfbox{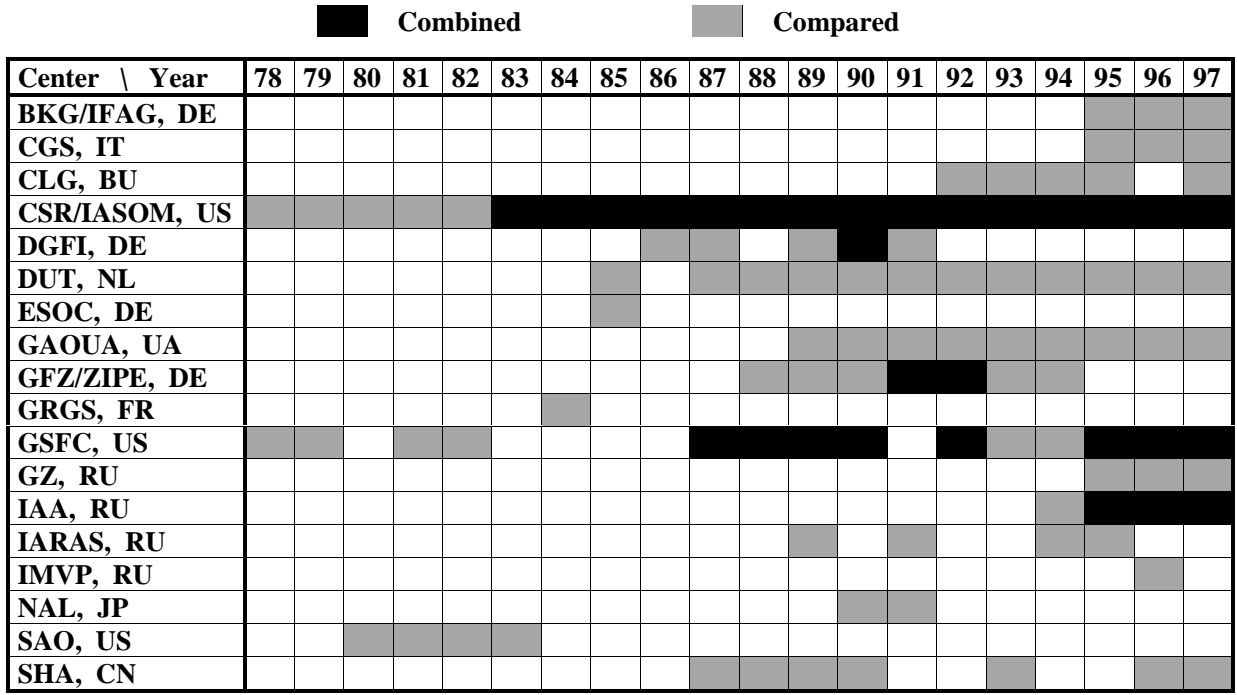}}
\caption{Contribution of SLR ACs to IERS yearly solutions.}
\label{fig:ac_contr}
\end{figure}

SLR observations of Lageos and ERP derived from these observations
are available from May 1976.
However, only in 1980--81 after significant improvement of range technique
SLR ERP series achieved accuracy required for investigation of Polar
motion (PM).
Figure~\ref{fig:ac_contr} summarizes information from IERS Annual Reports
for 1978--1997 concerning use of submitted SLR series in IERS yearly
solutions.
One can see that beginning from 1983 SLR ERP series is regularly used
for IERS combination.
It's also seen that the Center for Space Research, University of Texas
at Austin (CSR) provides most long-time spanned and stable PM series.

At present 5 analysis centers submit operational solutions
with 2-15 days delay, and about 10 analysis centers yearly
contribute final (up to 23 years) ERP series.
Most long series available for analysis are listed in Table
\ref{tab:long_series}.

\begin{table}
\parbox[t]{55mm}{
\caption{Longest series.}
\label{tab:long_series}
\def\arraystretch{0.78}
\begin{center}\begin{tabular}{ll}
CSR   & 1976 --- \\
GSFC  & 1980 --- \\
DUT   & 1983 --- \\
GAOUA & 1983 --- \\
IAA   & 1983 --- \\
CGS   & 1985 --- \\
BKG   & 1988 --- \\
\end{tabular}\end{center}
\caption{Agreement of series with C04, mas.}
\label{tab:acc_c04}
\def\arraystretch{0.78}
\begin{center}\begin{tabular}{lcc}
\multicolumn{1}{c}{Solution} & $X_p$ & $Y_p$ \\[3mm]
\hline
     &      &      \\
GPS  & 0.11 & 0.11 \\
SLR  & 0.11 & 0.12 \\
VLBI & 0.16 & 0.15 \\
\end{tabular}\end{center}
}
\hskip 5mm
\parbox[t]{70mm}{
\caption{Statistics of operational solutions based on USNO data for Sep 1998 -- Sep 1999.}
\label{tab:acc_usno}
\tabcolsep=4pt
\begin{center}\begin{tabular}{lrrc}
\multicolumn{1}{c}{Center} & \multicolumn{2}{c}{Delay, days} &
Error in PM, \\
\cline{2-3}
 & min & mean & mas \\
\hline
           &       &       &       \\
 NEOS VLBI &   6.3 & 12.0  &  0.10 \\
 IAA VLBI  &  13.3 & 22.5  &  0.12 \\[1em]
 CSR SLR   &   2.3 &  9.2  &  0.36 \\
 TUD SLR   &   4.6 &  7.8  &  0.36 \\
 IAA SLR   &   1.6 &  3.1  &  0.20 \\
 MCC SLR   &   7.6 & 11.1  &  0.21 \\[1em]
 IGS Final &  10.1 & 14.8  &  0.02 \\
 IGS Rapid &   1.1 &  1.4  &  0.06 \\
\end{tabular}\end{center}
}
\end{table}

\section{Accuracy of SLR PM series}

\begin{figure}
\centerline{\epsfbox{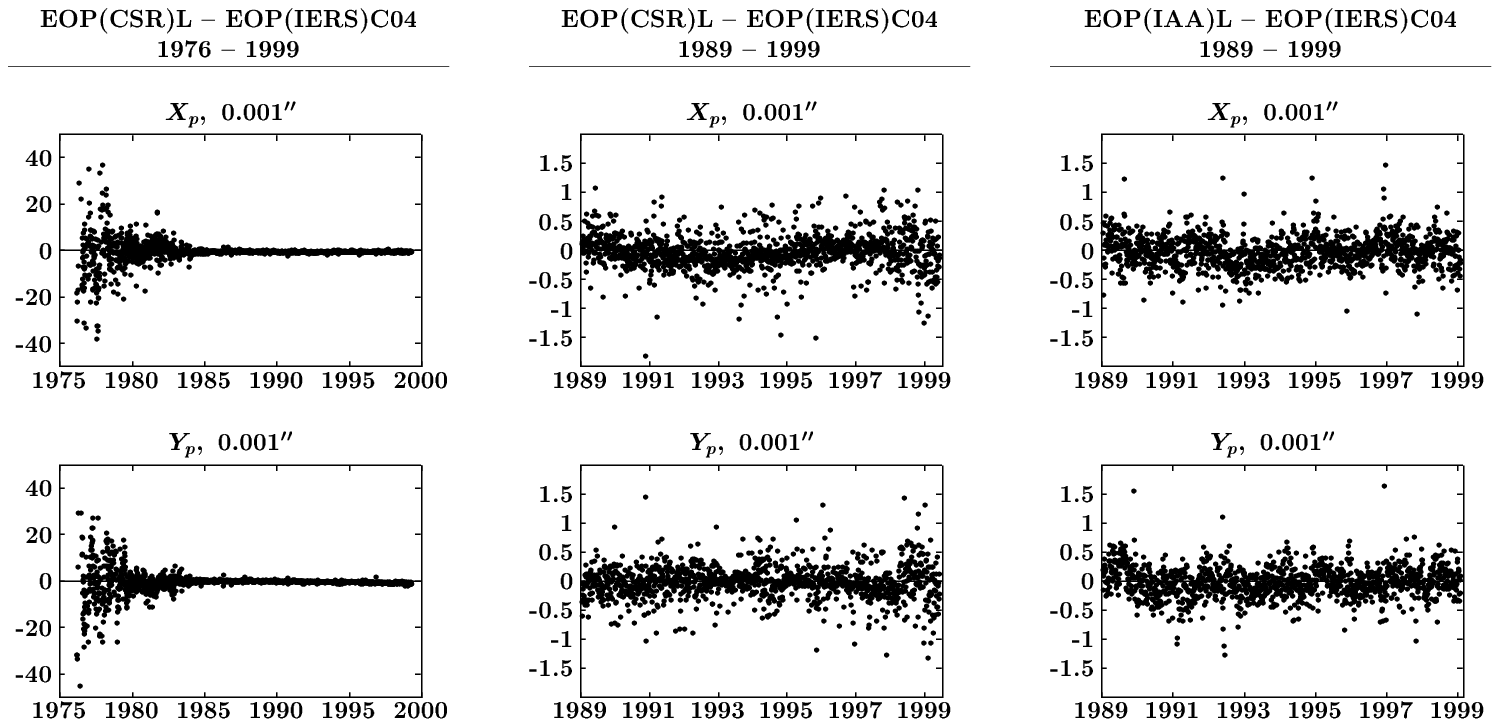}}
\caption{Differences between SLR and IERS PM series.}
\label{fig:csriaa}
\end{figure}

During long history of using SLR technique to study PM drastic
improvement in range precision was achieved.
It's interesting to see how this technology development affected
accuracy of PM series obtained from SLR observations.
Figure~\ref{fig:csriaa} shows differences between SLR series obtained
at the CSR and the Institute of Applied Astronomy, St.Petersburg (IAA)
and IERS combined solution EOP(IERS)C04.

One can clearly see in Figure~\ref{fig:csriaa} that replacement of
the first generation ranging equipment by the second generation
one about 1980 and its further replacement by the third generation
units leaded to significant improvement of accuracy of PM solution.
However, in spite of improvement of ranging technique and implementation
the fourth generation equipment, precision of PM series derived from
SLR observations remains the same during last 10--12 years.
We will try to discuss the problem below.

In 1998 IERS Central Bureau had derived final IERS combination
in two step: on the first stage combined VLBI, SLR, and GPS
series was computed and then these was used for final IERS combination.
Table~\ref{tab:acc_c04} copied from the 1997 IERS Annual Report
shows that precision of SLR and GPS combinations is practically
the same.  Unfortunately, this is not so for operational solutions.

Table~\ref{tab:acc_usno} compiled on basis of \verb"gpspol.asc"
files produced by USNO along with Bulletin A issues contains
statistics of series used for IERS Rapid Service.  Of course,
errors in PM obtained there depend on weighting applied to
combined series, but it already clear that accuracy of individual
SLR series is worse than VLBI and GPS ones, and delay of SLR
contribution is much worse than GPS one.
Nevertheless, one can see that at least two centers can produce
operational solution with delay about 2 days.
If all SLR analysis centers would provide such a solution on regular basis,
it will allow to have,
in principle, combined SLR solution
with accuracy 0.1--0.15 mas and delay 1.5--2 days.

\section{Factors limited quality of the SLR ERP solutions}

During long time SLR was the one of the main methods of determination
of PM and densification of PM series based on VLBI observations.
However, lately SLR ERP series are inferior to GPS ones
in quality (accuracy, density, delay of operational solution) of results.
It would be very important to understand existing problems in SLR observations
and analysis procedures and discuss possible ways to solve them.

\subsection{Observations}

The first problem with SLR observations is that
SLR technique is an 'one-object' one. This means that, unlike GPS,
SLR station can observe only one object in time. Hence, planning
of observations and priority
politics plays substantial role in acquisition of observations of
satellites on which PM determination is based, especially keeping
in mind that SLR observations are rather expensive and number of
ranging units are limited.

\begin{figure}[ht]
\epsfxsize=0.75\textwidth
\centerline{\epsfbox{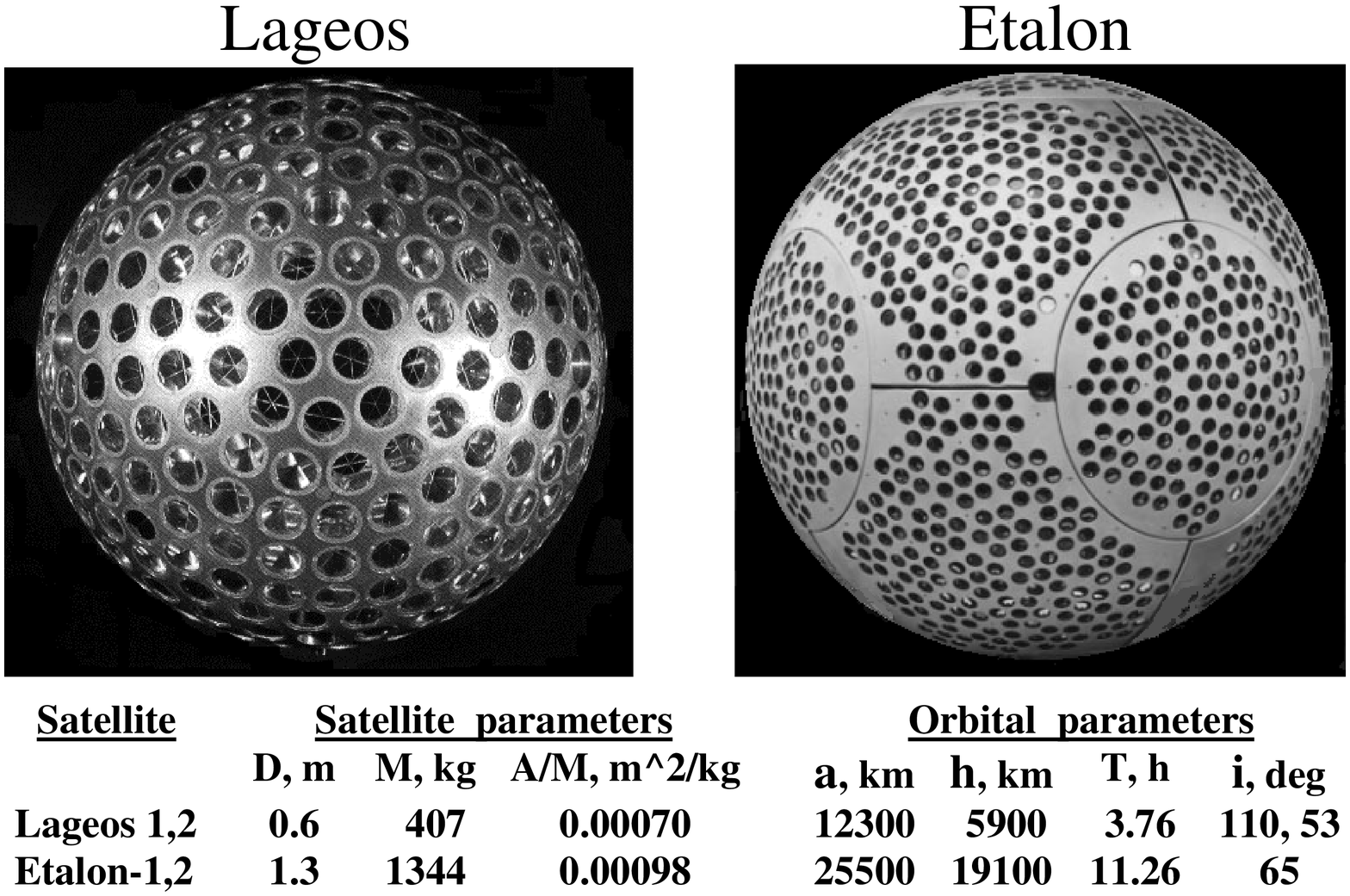}}
\caption{Satellites Lageos and Etalon.}
\label{fig:satellit}
\end{figure}

At present four operational satellites seem most suitable
for investigation of Earth rotation (as well as tectonic
motion and long-term temporal variations in various
geophysical parameters) --- two Lageos satellites and
two Etalon satellites.  Both Lageos and Etalon satellite was
launched to long-time stable orbits and have a low area to mass ratio.
Their description is presented in Figure~\ref{fig:satellit}.

\begin{figure}[ht]
\parbox{80mm}{
\epsfxsize=80mm
\epsfbox{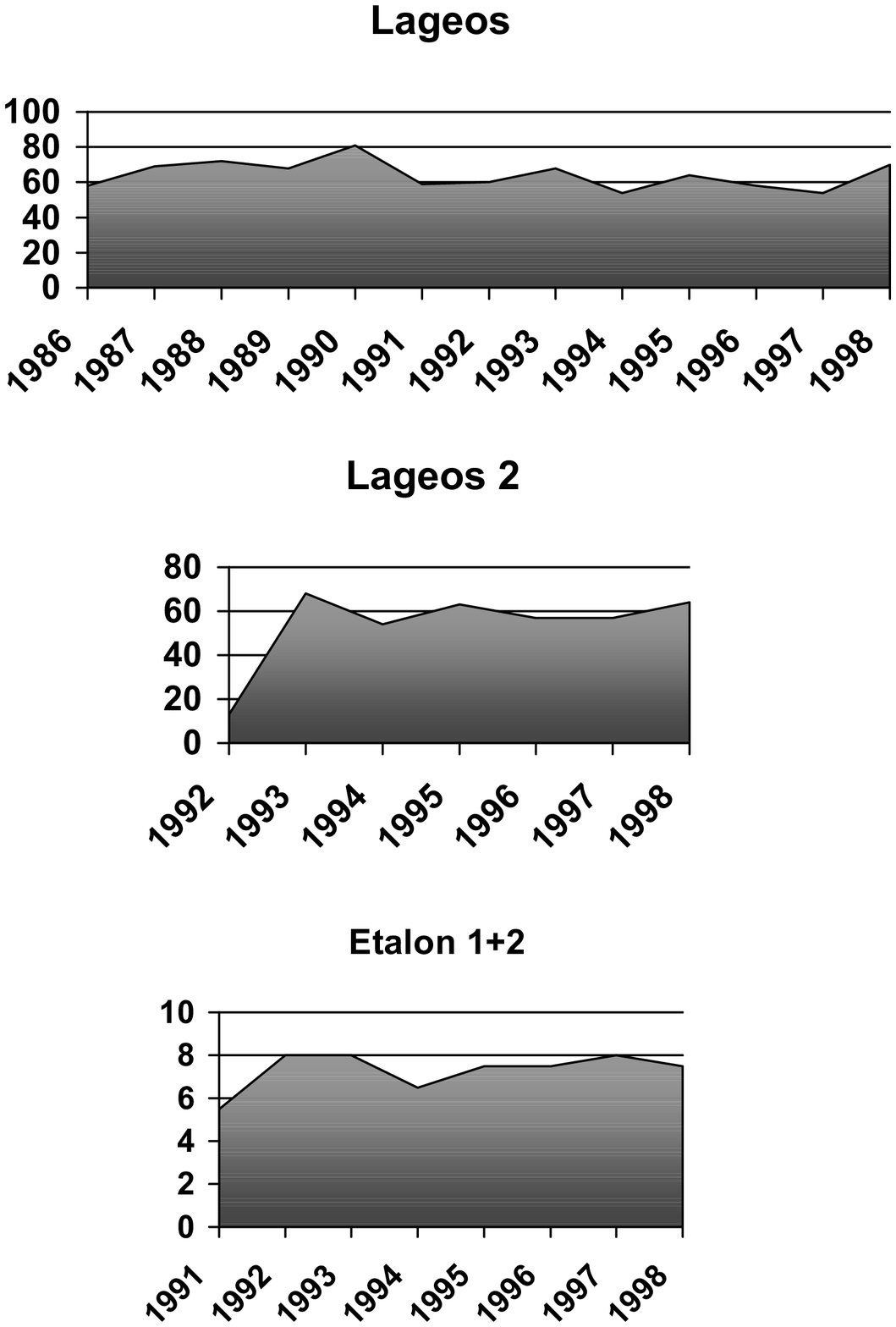}
%\caption{Number of observations (normal points) of satellites
%         Lageos and Etalon, thousands.}
%\label{fig:observ}
}
\hspace{15mm}
\parbox{80mm}{
%\vskip -30mm
\tabcolsep=4mm
\def\arraystretch{1.2}
%\caption{ILRS Priorities.}
%\label{tab:priorities}
\begin{tabular}{cl}
1 & ERS-1   \\
2 & GFO-1   \\
3 & ERS-2   \\
4 & Topex/Poseidon  \\
5 & GFZ-1       \\
6 & Geos-3      \\
7 & Starlette   \\
8 & WESTPAC     \\
9 & Stella      \\
10 & Ajisai     \\
{\large\bf 11} & {\large\bf Lageos 2}  \\
{\large\bf 12} & {\large\bf Lageos 1} \\
13-21 & GLONASS   \\
22-23 & GPS       \\
{\large\bf 24} & {\large\bf Etalon-1}  \\
{\large\bf 25} & {\large\bf Etalon-2}  \\
\end{tabular}
}
\caption{Number of observations (normal points), thousands (at the left)
         and ILRS priorities (at the right) of satellites.}
\label{fig:observ}
\end{figure}

Figure~\ref{fig:observ} shows number of observations of these satellites
during last 12 years and ILRS priorities.
One can see that number of observations of Lageos satellites remains
approximately the same during these years and number of observations
of Etalon satellite is too small to contribute seriously to analysis.

\begin{figure}
\epsfxsize=0.8\textwidth
\centerline{\epsfbox{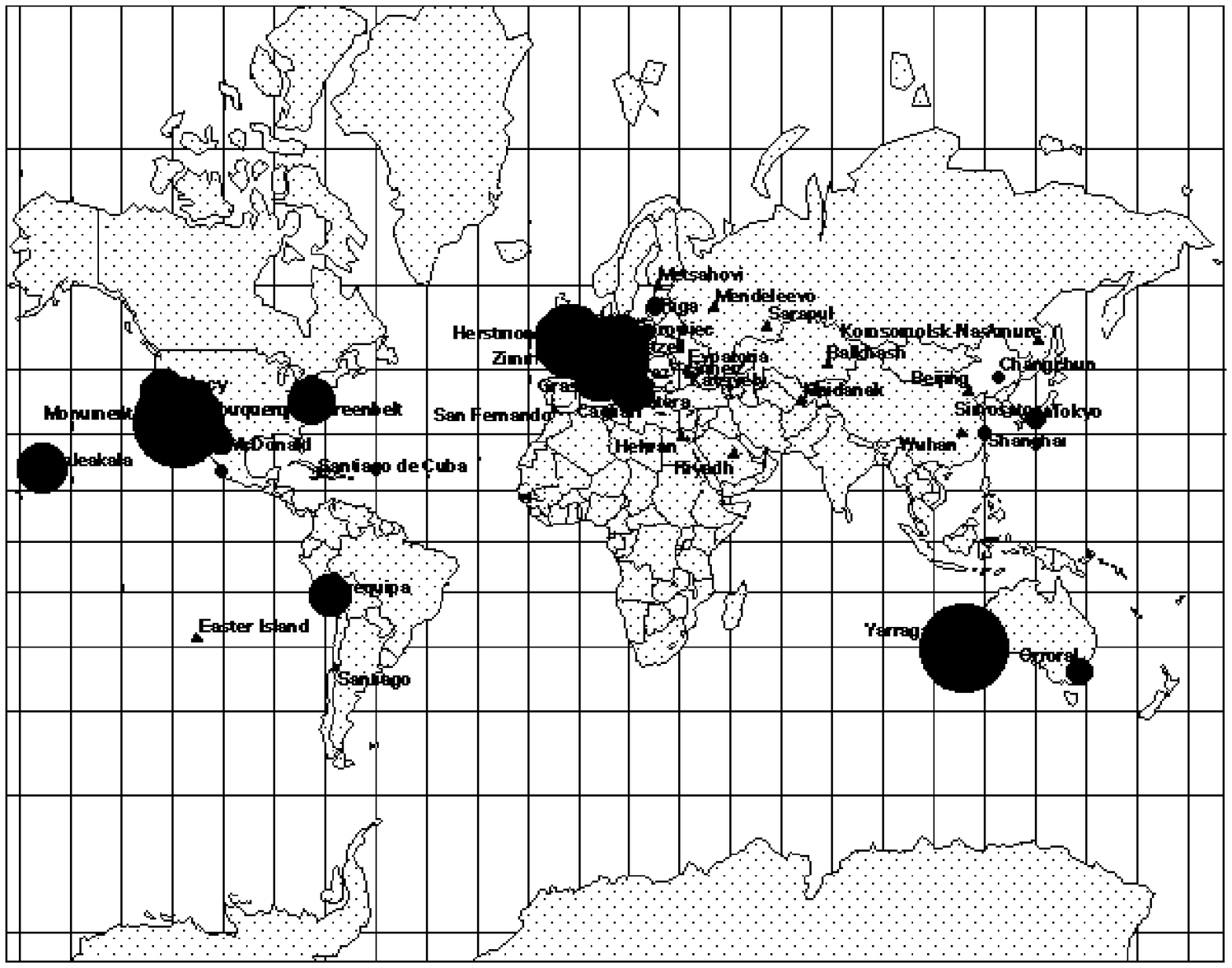}}
\caption{Distribution of observations in stations.}
\label{fig:sta_map}
\end{figure}

Let us see in more details distribution of SLR observations in stations
and time.
Figure~\ref{fig:sta_map} shows distribution
of observations in stations for the period from Sep 1983 till Aug 1999.
Table~\ref{tab:obs_sta} contains list of stations contributed
more than 2\% of total number of observations for whole period
and during the last year (in parenthesis).

\begin{table}
\caption{Distribution of observations in stations.}
\label{tab:obs_sta}
\small
\begin{center}\begin{tabular}{cllcll}
CDP & \multicolumn{1}{c}{Station} & \multicolumn{1}{c}{\%} &
CDP & \multicolumn{1}{c}{Station} & \multicolumn{1}{c}{\%} \\
\hline
7110 & MONUMENT PEAK & 9.6 (12.6) & 7843 & CANBERRA      & 3.0 (0.3) \\
7090 & YARRAGADEE    & 9.3 (10.0) & 7403 & AREQUIPA      & 2.7 (1.2) \\
7840 & HERSTMONCEUX  & 8.2 (8.7)  & 7810 & ZIMMERWALD    & 2.4 (3.6) \\
7210 & MAUI          & 5.4 (1.6)  & 7838 & SIMOSATO      & 2.4 (0.8) \\
7835 & GRASSE        & 5.3 (2.9)  & 7907 & AREQUIPA      & 2.0 (---) \\
7105 & WASHINGTON    & 5.3 (6.3)  & 7836 & POTSDAM       & 1.7 (2.1) \\
7839 & GRAZ          & 5.0 (7.9)  & 1884 & RIGA          & 1.5 (2.2) \\
7109 & QUINCY        & 4.8 (---)  & 7237 & CHANGCHUN     & 1.2 (2.5) \\
7939 & MATERA        & 4.6 (3.7)  & 7811 & BOROWIEC      & 1.0 (2.3) \\
8834 & WETTZELL      & 4.0 (4.8)  & 7849 & MOUNT STROMLO & 0.8 (7.9)\\
7080 & FORT DAVIS    & 3.3 (4.0)  & 7845 & GRASSE        & 0.6 (2.7)
\end{tabular}
\end{center}
Total about {\bf130} stations including mobile occupations \\[0.5em]
{\bf\phantom{1}5} stations provided {\bf38\%} of total number of observations \\
{\bf\phantom{1}7} stations provided {\bf48\%} of total number of observations \\
{\bf10} stations provided {\bf62\%} of total number of observations \\
{\bf15} stations provided {\bf75\%} of total number of observations \\
{\bf22} stations provided {\bf85\%} of total number of observations
\end{table}

The first impression is that stations
are distributed quite uniformly along of satellite orbit.
However, if we recall that SLR observations are weather-dependent,
it would be evident that Australia -- West Asia region need at least
one more active station.  Fortunately, several new stations was put
into operation in China and Japan during the last years but they are still
not active enough.
So, problem of more uniform distribution of observations in stations
evidently still exist.
Besides, as usually, most of stations are located in the
Northern hemisphere and at least two active stations would be
very desirable in the Southern hemisphere (in South America and
South Africa).

\begin{figure}[t]
\epsfxsize=0.40\textwidth \epsfbox{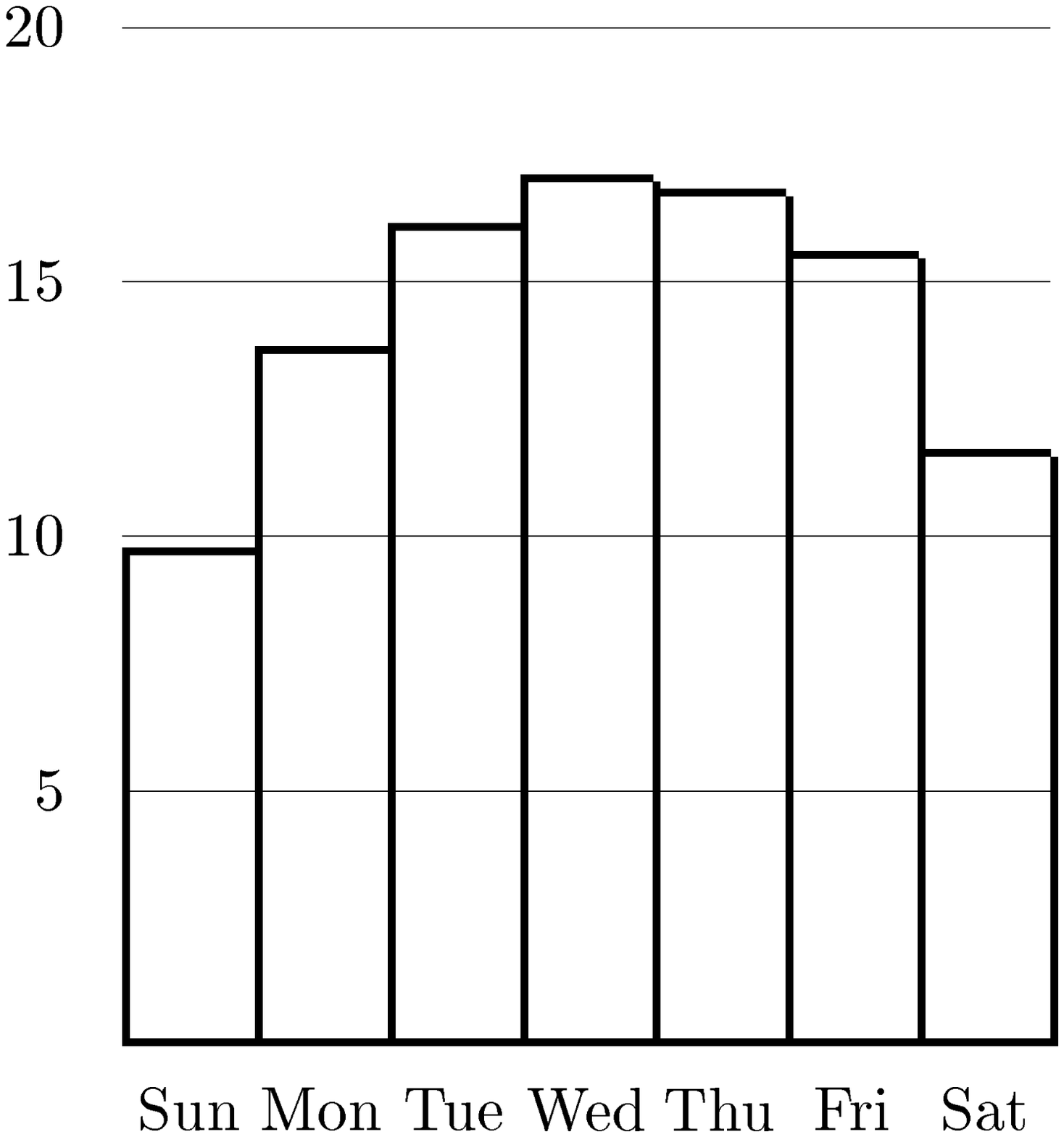}
\hspace{0.05\textwidth}
\epsfxsize=0.55\textwidth \epsfbox{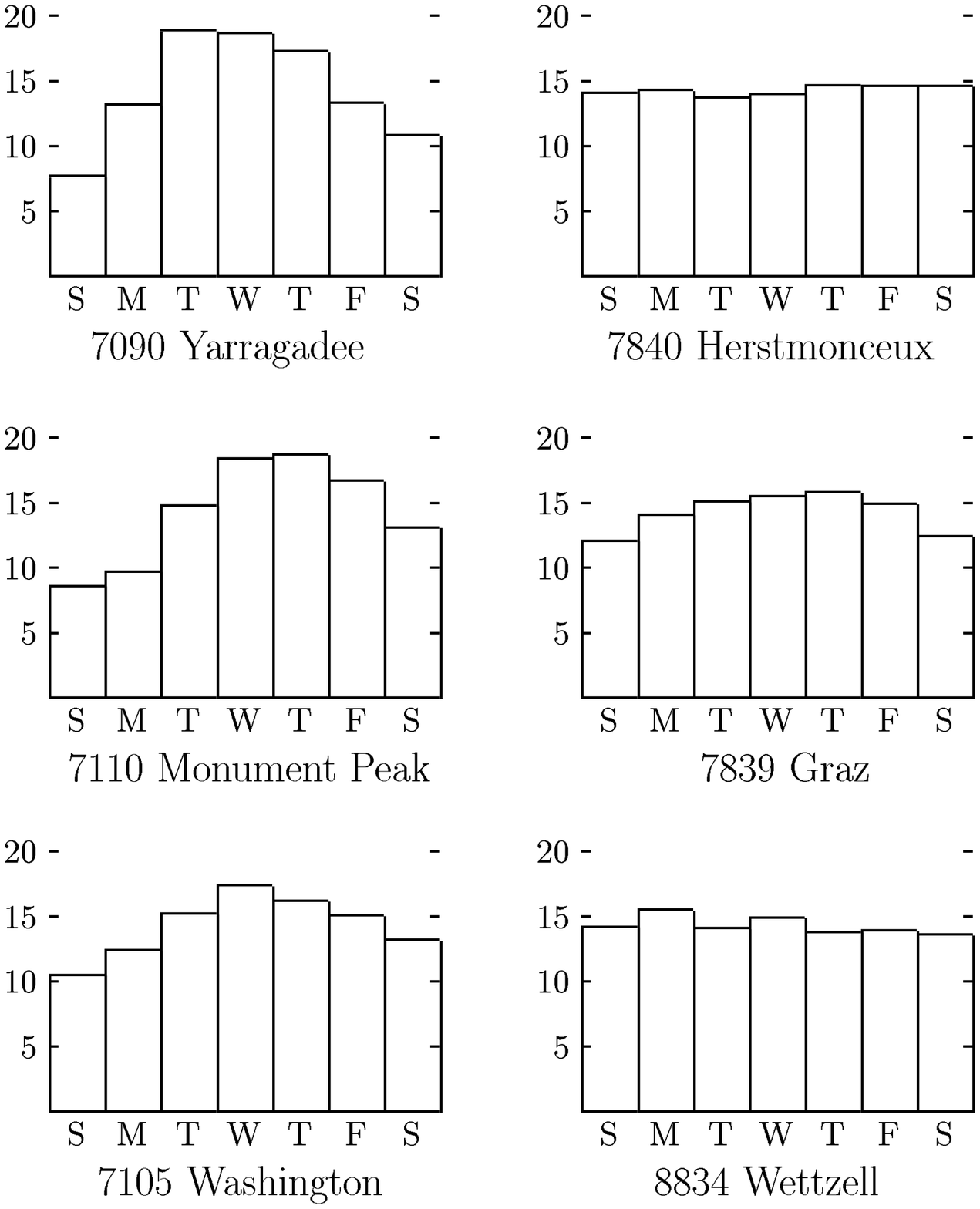}
\caption{Distribution of observations in days of week for all stations (left) and selected stations (right).}
\bigskip
\label{fig:obs_week}
\end{figure}

Distribution of observations in months is more or less uniform,
with small decrease in November and December (sharp dip in Christmas days!).
But distribution in days of week (Figure~\ref{fig:obs_week}) is
not so good.  One can see that number of observations on several
most active stations is about twice lesser than during workdays.
This is statistics for whole interval from Sep 1983 till Aug 1999.
For the last year distribution is a little bit more uniform -
maximum number of observations was made in Wednesday (16.2\%),
minimum in Sunday (11.6\%), but still far from ideal.

\subsection{Analysis}
\label{sect:analysis}

Though, in my opinion, the only real way to improve quality of
SLR ERP series is to increase number of observations and, especially,
number of satellites involved in determination of ERP, analysis
strategy can also be advanced.

Modern tendency is to use combined VLBI, GPS, SLR series for final
IERS combination.  Unfortunately, PM series
produced by SLR analysis centers are not unified enough
which makes difficult their comparison and combination.
The main problems in individual SLR series are:
\begin{itemize}
\item Sparse time series (3-5 days for most solutions, 1 day needed).
\item Large delay of operational solutions (5-10 days for most solutions,
     1-2 days needed).
\item Non-unified TRS.
\item Dependence on a priory values of EOP.
\end{itemize}

Analysis of methods used for determination of ERP from SLR observations
shows that not all possibilities for improvement and unification of SLR ERP
series are exhausted.  What can we do to make SLR series more efficient?
Existing experience shows that significant improvement could be achieved
if we would:
\begin{itemize}
\item Produce daily EOP series.
\item Produce operative solution with delay about 2 days.
\item Advance strategy for determination UT.
\item Use the same TRS for all solutions (ITRF seems most suitable for
      this purpose).
\end{itemize}

An analysis of these problems and some proposals
on their solution was presented in (Malkin 1998).
Below proposed methods tested in the IAA during last three years
are briefly described.

The first serious problem is how to compute daily ERP series without
loss of accuracy. There are two main ways to compute daily series.
The first is to use daily arc solutions (i.e. use one day arcs).
But this leads to degradation of accuracy of solution.
The second one is to use overlapping arcs with appropriate length
and one day shift between arcs.
Such a solution produces dependent results for neighbor arcs,
but significant improvement in precision can be considered as
sufficient compensation.
Table~\ref{tab:arc_acc} shows accuracy of ERP solution with various
length of arc.

\begin{table}
\caption{Dependence of errors in ERP on length of arc
(RMS of EOP(IAA)L -- EOP(IERS)C04 after removing trend).}
\begin{center}
\tabcolsep=1em
\begin{tabular}{lccccc}
\raisebox{-1.6ex}[0pt][0pt]{ERP} & \multicolumn{5}{c}{Length of arc, days} \\
\cline{2-6}
& 1 & 3 & 4 & 5 & 7 \\
\hline
$X, 0.001''$     & 0.40 & 0.23 & 0.20 & 0.18 & 0.16 \\
$Y, 0.001''$     & 0.34 & 0.22 & 0.21 & 0.18 & 0.15 \\
$LOD, 0.0001 s$  & 0.44 & 0.16 & 0.15 & 0.12 & 0.09 \\
\end{tabular}
\end{center}
\label{tab:arc_acc}
\end{table}

So, experience shows that using overlapping arcs yields more precise
ERP series than using one day arcs.

It would be very desirable to have strictly daily series for all analysis
centers,
i.e. series with epochs equal to $0^h$.  This can be achieved by
two methods - determination of Pole coordinates with their rates
and a posteriori interpolation to $0^h$.  Numerous test showed that
including Pole coordinates rates into parameter model lead to
small degradation of accuracy of PM, evidently due to degradation
of normal system matrix condition.
A posteriori interpolation of ERP to $0^h$ epochs gives better
result. Linear interpolation was found as most suitable for this purpose.

Table~\ref{tab:daily_int} shows dependence of
results on effect of interpolation of raw ERP series.
One can see that use of linear interpolation don't lead to
visible degradation of accuracy.  Special tests showed that
strictly daily series obtained with interpolation is more
accurate that series obtained with including Pole coordinates
rates in parameterization.

\begin{table}
\caption{Effect of interpolation of SLR ERP series
(RMS of EOP(IAA)L -- EOP(IERS)C04 after removing trend).}
\begin{center}
\tabcolsep=1em
\begin{tabular}{lcc}
\raisebox{-1.6ex}[0pt][0pt]{EOP} & \multicolumn{2}{c}{Series} \\
\cline{2-3}
& ~~~~~Raw~~~~~~ & Interpolated \\
\hline
$X, 0.001''$     & 0.154 & 0.155 \\
$Y, 0.001''$     & 0.171 & 0.171 \\
$LOD, 0.0001 s$  & 0.154 & 0.168 \\
\end{tabular}
\end{center}
\label{tab:daily_int}
\end{table}

It should be mentioned also that
producing daily solution (independently which method id used) provide
operational solution with steady delay about 2 days, which solves the
second problem mentioned above.

Other serious problem is determination of UT from SLR (and other
satellite observations).
It is well known that UT1 cannot be separated from longitude of node
of satellite orbit during parameter solution.
Three methods are being used to solve this problem:
\begin{itemize}
\item Fixing longitude of node during parameter solution (usually,
during last iteration).
\item Analysis of node longitude series, forecasting it and use
predicted values for operational solution.
\item Integrating LOD series to obtain independent free-running
UT series with its possible correction for high-frequency variations from
comparison with VLBI series.
\end{itemize}

Evidently, only the latter method can provide (in principle) independent
result.
Since that is not a subject of this paper, we will not stay
on detailed analysis of this problem.
However, It is worth to mention here that significant improvement of
SLR UT series is also impossible without increasing of number of
satellites involved in determination of ERP.

At last, use the same terrestrial reference frame
for all SLR solutions seems evident to
achieve uniform solutions for combination. Use of ITRF as
terrestrial reference frame for by all analysis centers
for their SLR solutions provides more homogeneous series
for SLR combined solution.

Realization
of this or alternative analysis strategy could provide more uniform,
accurate and operative SLR ERP series. After that, combining of all submitted
series to final ILRS SLR product seems reasonable and useful for further use
for IERS and other purposes.

We have not mention here such a serious problem as dependence of
SLR ERP results on a priori values.  This is worth to perform
special investigation for each method used for computation
of ERP at various analysis centers.

\section{Conclusion}

Satellite laser ranging technique made and make a very valuable contribution
to Earth dynamics.  In particular, very valuable contribution was made
in investigation of PM.  During many years SLR was one of the main
methods of determination of polar motion and main method of densification
of ERP series obtained with VLBI.
At present 5 analysis centers submit operational solutions
with 2-15 days delay, and about 10 analysis centers yearly
contribute final (up to 23 years) ERP series.

However, due to principal peculiarity of this method
(relatively expensive experiment, limited number of units, lack
of capability of multi-satellite ranging, etc.)
quality of SLR ERP data remains
the same during the last decade in spite of ranging precision
improved by a factor of a thousand from a few meters to a few millimeters
since the first SLR experiments (and by factor of about 10 during last
10-12 years).
This leads to decreasing of weight of SLR solutions in the combined
IERS EOP series.

It is evident that only substantial increasing of number of observations
and satellites involved in investigation of Earth rotation and improvement
of distribution of observations in stations and time can help
in improvement of the SLR ERP accuracy.
However, capacity of tracking network is practically exhausted.

On the other hand, in spite of GPS provides determination of ERP
with impressive accuracy and delay, SLR results are very
important for combined IERS solution for improvement of systematic
accuracy of the final IERS product.
Analysis of precision of individual SLR series
shows that its combination can provide combined SLR much more accurate and
rapid series.
To achieve highest
accuracy of combined SLR ERP product is necessary to solve problems
discussed in section \ref{sect:analysis}.

Realization of this opportunities by ILRS would be very important
for investigation of Earth rotation because allow to save independent
method of determination of PM and velocity of the Earth rotation.

\section*{Acknowledgments}
Author is very grateful to Scientific Organizing Committee for invitation
to the meeting and financial support of this trip.

\bigskip
\noindent{\Large\bf References}
\bigskip
\leftskip=\parindent
\parindent=-\leftskip

Malkin, Z. 1998, On unification and combination of SLR EOP series, Presented at the 1998 IERS Workshop, Potsdam, Germany, Sep 28 -- Oct 2.

Tapley,~ B.~D., Schutz,~B.~E., Eanes,~ R~.J., Ries,~J.~C., Watkins,~ M~.M. 1993, in Contribution of Space Geodesy to Geodynamics:
  Earth Dynamics, D.~Smith \& D.~Turcotte, Washington: AGU, 147.

\end{document}